\documentclass[preprint,12pt,showkeys,showpacs]{revtex4}
\usepackage{amsmath}
\usepackage{latexsym}
\usepackage{amssymb}
\usepackage{graphics,epstopdf}
\usepackage{amsmath,amssymb,amsthm}
\usepackage{slashed}
\usepackage[colorlinks=true, citecolor=blue, urlcolor=blue ]{hyperref}

\begin{document}
\title{Current conservation in  charge conjugation parity time reversal symmetry (CPT) violating gauge-invariant nonlocal Thirring model}

\author{P Patra}
\thanks{corresponding author}
\email{monk.ju@gmail.com}
\affiliation{Department of Physics, University of Kalyani, West Bengal, India-741235}

\author{J P Saha}
\email{jyotiprasadsaha@gmail.com}
\affiliation{Department of Physics, University of Kalyani, West Bengal, India-741235}

\begin{abstract}
Charge conjugation, parity transformation and time reversal symmetry   (CPT) violation and Lorentz invariance can coexist in the framework of non-local field
 theory. In this article we have proposed a class of Charge conjugation, parity transformation and time reversal symmetry (CPT)
 violating Lorentz invariant nonlocal gauge-invariant models, which can be termed as non-local Thirring models.  The conserved currents in this aspect are obtained. 
\end{abstract}

\maketitle
\section{Introduction}
 Within Lorentz invariant framework, every local quantum field theory obey the well established CPT theorem. This is immediately followed by the equality
 of mass of particle and corresponding antiparticle. Therefore, it is natural to give the first effort by using the non-local theory to deal with the
 particle-antiparticle mass splitting phenomenon which has been speculated by recent data analysis \cite{1,2,3,4}. 
 The concrete mathematical formulation as well as physical implications of non-local field theory have been studied in great detail in literature of 
theoretical  physics by several authors \cite{5,6,7,8,9,10,11,12,13,14}. 
\\

 The meaning of particle and corresponding anti-particle is apparently not so clear and it is not well understood whether such nonlocal theory
 can have sensible S-matrix. But, if we adopt the usual meaning of particle and corresponding anti-particle from the idea of local field theoretical models,
 one can see that inclusion of non-locality may break the equality of mass of particle and antiparticle.
 An effort to explain the particle anti-particle mass splitting is given in  \cite{15,16,17,18} in the framework of CPT violating, but Lorentz invariant, 
 non-local interactions. 
\\

 To keep track with gauge-invariance in non-local theory, one can easily understand that we have to introduce 
non-integrable phase factor (this can be replaced by a first quantized very massive particle propagation; an analogue of the
 Chan-Paton factor in string theory \cite{18}) in nonlocal interaction part. It is a successful
 attempt by Chaichian et al \cite{18} to incorporate the gauge-invariant scenario in Lorentz invariant CPT-violating
 non-local cases. For an introduction to the non-local field theory, one can see Ref. \cite{19}. 
\\ 

The models discussed  by Chaichian et al. \cite{18} are in $1+3$ dimension. But, all the formalisms of \cite{15,16,17,18,19} can
 directly be used in $1+1$ dimensional case. 
As one of our motivation is to study the Thirring model, so we confine ourselves in $1+1$
dimension.  Thirring model is a completely soluble $1+1$- dimensional quantum field theory covariant with respect to Lorentz transformations in $1+1$ 
dimensions \cite{20,21,22,23,24}.  One of the important observation is that
the fundamental fermion of the Thirring model can be identified with the soliton of the $Sine$- Gordon 
 \cite{20, 25,26} equation which is the theory of massless scalar field.

\section{Gauge invariant Nonlocal Thirring models}
\subsection{Models under Consideration}
We consider the action
\begin{eqnarray}
S=\int d^2x \{\bar{\psi}(x)i \slashed{D}\psi(x)-m\bar{\psi}(x)\psi(x)- 
\lambda j^\mu(x)j_\mu(x) - \nonumber \\  \tilde{\mu} \int d^2y \theta(x^0-y^0)
\delta((x-y)^2 -l^2)J(x,y) + \mbox{C.C}\} -  \nonumber \\  \frac{1}{4}\int d^2x F^{\mu\nu}(x)F_{\mu\nu}(x)
\end{eqnarray}

Here, $\psi$ are the Dirac fields, $\slashed{D}=\gamma^\mu D_\mu$, $\gamma^\mu$ are the usual Dirac matrices, $D_\mu=\partial_\mu-ie A_\mu$ is the
 covariant derivative, $j^\mu(x)=\bar{\psi}(x)\gamma^\mu\psi(x)$, $\lambda$ and $\tilde{\mu}$ are the coupling constants, $\theta(x)$ stands for the usual
 Heviside step function, $\delta(x)$ is the usual Dirac delta function, $F^{\mu\nu}(x) = \partial_\mu A_\nu- \partial_\nu A_\mu$. $J(x,y)$ in the nonlocal
 term which will be different for different cases under consideration. The coupling constant $\tilde{\mu}$ will be different for different cases. 
\\
In general, inclusion of non-locality may break the local gauge-invariance of the theory. The problem can be solved \cite{18} by introducing
 Schwinger's non-integrable phase factor $e^{ie\int_{y}^{x}A_\mu(z)dz^\mu}$ in the interaction term.\\ 
In this article we have considered the following types of interaction 
\\

{\bf Case-I} 
\begin{eqnarray}
S_1^{int}=-i\mu \int d^2 x d^2y \bar{\psi}(x)\gamma^\mu e^{ie\int_y^x A_\mu (z)dz^\mu}\psi(y) \theta(x^0-y^0) \nonumber \\ \delta((x-y)^2-l^2)j_\mu(y) +  C.C
\end{eqnarray}
Here, we have chosen $\tilde{\mu}=i\mu$ (where, $\mu$ real) purely imaginary. 
The action is invariant under the gauge-transformation 
\begin{eqnarray}
\psi(x)\rightarrow e^{i\alpha(x)}\psi(x) \nonumber \\
A_\mu(x)\rightarrow A_\mu(x)+\frac{1}{e}\partial_\mu \alpha(x)
\end{eqnarray} 
The other  interaction which is invariant under the same gauge transformation reads\\
{\bf Case-II} 
\begin{eqnarray}
S_2^{int}=-\mu \int d^2 x d^2y \bar{\psi}(x)\gamma^\mu e^{ie\int_y^x A_\mu (z)dz^\mu}\psi(y) 
(\theta(x^0-y^0)-\theta(y^0-x^0)) \nonumber \\\delta((x-y)^2-l^2)\bar{\psi}(x)  e^{ie\int_y^x A_\mu (z)dz^\mu}\gamma_\mu \psi(y)
\end{eqnarray}

Here, $\tilde{\mu}=\mu$ is real. 

\subsection{CPT Non-conservation}

The combination of discrete transformation of charge conjugation, parity and time reversal (CPT), effectively flips the
 sign of all coordinates and also performs a complex conjugation. By applying the CPT operation in our models, one can easily see that CPT is not conserve
 in all the above mentioned nonlocal interactions. \\
 In particular, for Case-I, the nonlocal interaction term transforms under CPT as
\begin{eqnarray}
-i\mu \int d^2 x d^2y \bar{\psi}(y)\gamma^\mu e^{ie\int_y^x A_\mu (z)dz^\mu}\psi(x)
  \theta(x^0-y^0)\delta((x-y)^2-l^2)j_\mu(x) \nonumber \\ + i\mu \int d^2 x d^2y j^\mu(y) \theta(y^0-x^0)\delta((x-y)^2 -l^2) 
 \bar{\psi}(y)\gamma_\mu e^{ie\int_y^x A_\mu (z)dz^\mu}\psi(x) 
\end{eqnarray}
which is not identical with the actual interaction term. \\
Also, by direct applying the CPT operation, it is not difficult to see that, the second type of interaction term (Case-II) transform as
\begin{eqnarray}
\mu \int d^2 x d^2y \bar{\psi}(x)\gamma^\mu e^{ie\int_y^x A_\mu (z)dz^\mu}\psi(y) [\theta(x^0-y^0) - \theta(y^0-x^0)]\delta((x-y)^2-l^2) \nonumber \\ 
\bar{\psi}(x) e^{ie\int_y^x A_\mu (z)dz^\mu}\gamma_\mu\psi(y) 
\end{eqnarray}
i.e, $CPT=-1$.\\
One can see that if $\mu$ is purely imaginary, CPT is conserved in that case. \\

It is interesting that the CPT- and T-violating term is real in the present case. T-violating terms usually carry imaginary coupling constants in the
ordinary local field theory.
\\
In the next section it is shown how the conserved current looks like for these non-local models.

\section{Current conservation}
Using proposal of \cite{18}, we can replace non-integrable phase factor (considering the non-integrable phase factor
 as an independent dynamical entity) by a
first quantized very massive particle propagation defined by covariant path integral
\begin{eqnarray}
e^{ie\int_{y}^{x}A_\mu(z)dz^\mu}\delta_{\alpha,\beta} \Rightarrow \int \mathcal{D}z^\mu \exp\{i 
\int_{y}^{x} \frac{1}{2}[(\dot{z}^\mu)^2  +M^2]d\tau \nonumber \\ + ie \int_{y}^{x}A_\mu(z) \frac{dz^\mu}{d\tau} d\tau\} \delta_{\alpha,\beta}
\end{eqnarray}
The consequences of that can be seen from \cite{18} and \cite{27}.
\\
As the fermion pair creation can be examined through the lowest order interaction, we expand the non-integrable phase factor
 to the lowest order in $\it{O}(e)$ as follows
\begin{equation}
e^{ie\int_{y}^{x}A_\mu(z)dz^\mu}=1+ie\int_{y}^{x}A_\mu(z)dz^\mu
\end{equation}
Now we shall study each of the two cases separately. 

\subsection{Case- I} 
The electromagnetic current is defined by \cite{18}
\begin{equation}
\mathcal{J}^\mu (w)=\frac{\delta}{\delta A_\mu(w)}S^I 
\end{equation}
Where $S^I$ is interaction term.\\
Interaction part for the lowest order pair creation is given by
\begin{eqnarray}
S_1^I=e\int d^2x\bar{\psi}(x)\slashed{A}(x)\psi(x) +   e \mu \{\int d^2x d^2y  \bar{\psi}(x)
\gamma^\mu [\int_{y}^{x} A_\mu (z) dz^\mu  \nonumber \\ \psi(y) \Theta(x^0-y^0)  \delta((x-y)^2-l^2) j_\mu(y) ] + C.C \}
\end{eqnarray}
Where $z^\mu$ stands for the coordinate of the massive particle. 
\\
And the corresponding electromagnetic current is
\begin{eqnarray}
\mathcal{J}_1^\mu (w)=\frac{\delta}{\delta A_\mu(w)}S_1^I \nonumber \\
=  e\bar{\psi}(w)\gamma^\mu\psi(w) - e\mu \int d^2y \bar{\psi}(w)\gamma^\mu\psi(y)  \theta(w^0-y^0)\nonumber \\ \delta((w-y)^2
-l^2)j_\mu(y)+ e\mu \int d^2y \bar{\psi}(y)\gamma^\mu\psi(w) \nonumber \\ \theta(y^0-w^0) \delta((w-y)^2-l^2)j_\mu(w) - C.C
\end{eqnarray}
Here, we have used 
\begin{equation}
\frac{\partial}{\partial w^\mu}\delta(z(\tau)-w)\frac{dz^\mu}{d\tau} = -\frac{d}{d\tau}\delta(z(\tau)-w)
\end{equation}
 In true sense the inclusion of path integral for the very massive particle  will alter the $\delta (z(\tau)-w)\frac{dz^\mu}{d\tau}$ term as 
\begin{eqnarray}
\frac{1}{Z}\int d^2 z'\delta^2(z'(\tau)-w)\langle x,\tau_x|z',\tau\rangle\frac{dz'^\mu}{d\tau}\langle z',\tau|y,\tau_y\rangle \nonumber \\
=\frac{1}{Z}\langle x,\tau_x|\delta^2(z(\tau)-w)\frac{dz\mu(\tau)}{d\tau}|y,\tau_y\rangle
\end{eqnarray}
where $Z$ is the normalization factor of the path integral partition function. \\ 
Now the current $\mathcal{J}_1^\mu (w)$ is indeed conserve $(\partial_\mu \mathcal{J}_1^\mu (w)=0)$ by virtue of the equation of motion
 and the conjugate equation. 

\subsection{Case- II} 
 The relevant part for the lowest order pair creation in that case is given by
\begin{eqnarray}
S_2^I=e \int d^2x  \bar{\psi}(x)\slashed{A}(x) \psi(x) - 2i e\mu \int d^2x d^2y    \bar{\psi}(x)\gamma^\mu  [\theta(x^0
-y^0)- \nonumber \\ \theta(y^0-x^0)]  \delta((x-y)^2-l^2)(\int_y^x A_\mu (z)dz^\mu)\psi(y)\bar{\psi}(x)\gamma_\mu\psi(y)
\end{eqnarray}
And the current,
\begin{eqnarray}
\mathcal{J}_2^\mu(w)= e \bar{\psi}(w) \gamma^\mu \psi(w) -i e\mu\int d^2 y [\theta(y^0-w^0)-\theta(w^0-y^0)] \nonumber \\ \delta((y-w)^2
-l^2)  \bar{\psi}(y)\gamma^\mu\psi(w)\bar{\psi}(y)\gamma_\mu\psi(w) + C.C
\end{eqnarray}
which is conserved by virtue of the equation of motion and conjugate equation.

\section{Results and Discussion}
We have proposed a class of CPT violating nonlocal models which can be termed as nonlocal Thirring models.
 Our model is invariant under $\it{U}(1)$ gauge transformation. 
The modified currents which are conserved are given explicitly in eqs. (11) and (15).

\section{Conclusions}
We have seen that for non-local theory, with the help of inclusion of Swinger 
non-integrable phase factor, we can make the theory gauge invariant, which gives us the conserved currents. 
\\
Intuitively one may argue that inclusion of intermediate very massive (indefinite) particle ( eq. (7)) propagation may break the equality
 of masses of particle and corresponding anti-particle and one will observe the possible mass-splitting between particle and antiparticle.
 One can see that particle antiparticle mass-splitting is evident for the discussed cases given in \cite{28}. However, 
particle anti-particle mass-splitting mentioned in Ref. \cite{1,2,3} and interpreted in \cite{15,16,17,18,19} takes place in four dimensions,
 while our study is in $1+1$ dimension. 
We can say that, the idea of four dimensions may be used in the lower dimensional field theory as well.
\\ 
In general, all known  nonlinear integrable systems have soliton like solution. Non-locality may break the profile of solitonic nature. 
This may be an interesting study.

\section{Acknowledgement}
 P Patra is grateful to CSIR, Govt. of India for fellowship support. J P Saha likes to acknowledge the 
financial support from DST-PURSE (Grant No:FD/1626,2013), Govt. of India. We would like to thank Masud Chaichian for several suggestions which were 
helpful for this work.

\end{document}